\documentclass[aps,prl,twocolumn,showpacs,superscriptaddress,groupedaddress,floatfix]{revtex4}
\usepackage[english]{babel}
\usepackage{amsmath,amsthm,amssymb}
\usepackage{amsfonts}
\usepackage{amsopn}
\usepackage{float}
\usepackage{setspace}
\usepackage{fancybox}
\usepackage{listings}
\usepackage{url}
\usepackage{appendix}
\usepackage{dsfont}
\usepackage{graphicx}
\usepackage{dcolumn}
\usepackage{bm}
\usepackage{color}
\begin{document}

\widetext

\title{Shot-Noise in Fractional Wires: a Universal Fano-Factor Different than the Tunneling Charge}

\author{Eyal Cornfeld}
\author{Izhar Neder}
\author{Eran Sela}
\affiliation{Raymond and Beverly Sackler School of Physics and Astronomy, Tel-Aviv University, Tel Aviv, 69978, Israel}

\begin{abstract}
We consider partially gapped one dimensional (1D) conductors connected to normal leads, as realized in fractional helical wires. At certain electron densities, some distinct charge mode develops a gap due to electron interactions, leading to a fractional conductance. For this state we study the current noise caused by tunneling events inside the wire. We find that the noise's Fano-factor is different from the tunneling charge. This fact arises from charge scattering at the wire-leads interfaces. The resulting noise is, however, universal - it depends only on the identification of the gapped mode, and is insensitive to additional interactions in the wire. We further show that the tunneling charge can be deduced from the finite frequency noise, and yet is interaction dependent due to screening effects.
\end{abstract}

\pacs{73.23.-b, 71.10.Pm, 71.70.Ej}
\maketitle

\paragraph{Introduction \--}
Shot noise appears in the current fluctuations and provides valuable information on electron-electron correlation effects in mesoscopic systems~\cite{Blanter}. In particular its interpretation as a measurement of the unit electric charge that tunnels through a barrier is of great importance. It allowed the celebrated direct measurement of the fractional charge of Laughlin quasiparticles (QP) in fractional quantum Hall (FQH) systems~\cite{Picciotto}. It even remains so far the main experimental handle on predicted non-abelian FQH states \emph{e.g.} in the $\nu=\frac{5}{2}$ Moore-Read state~\cite{Dolev}.

An ideal situation for such a direct interpretation of shot-noise $S$
occurs when a current $I_{\rm{tun}}$ consists of rare tunneling events of quasiparticles between two channels in a strongly correlated electron system.
In such general circumstances the Fano-factor, given by the ratio $q^*=\frac{S}{2 I_{\rm{tun}}}$,
matches the total charge transferred per tunneling event between the leads. Usually this charge corresponds to the local elementary excitation charge defined \emph{inside} the strongly correlated system (\emph{i.e.} the QP charge).

\begin{figure}
\includegraphics[width=\linewidth]{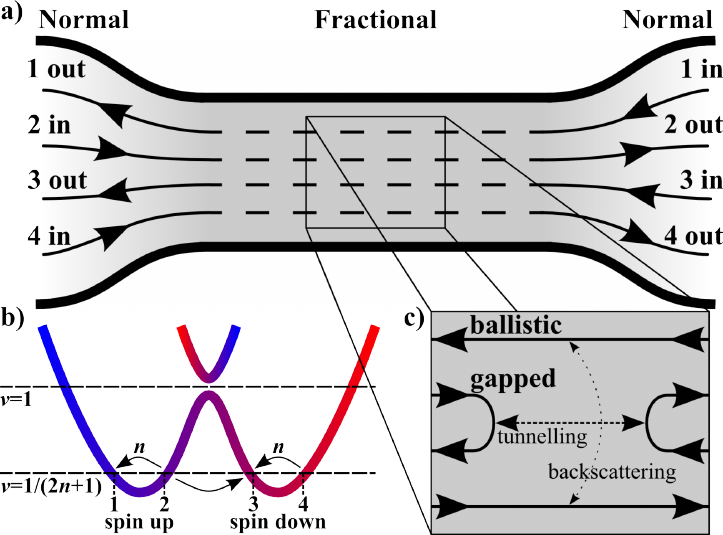}
\caption{\label{fg:1} a) The chiral currents, incoming into the fractional wire from the normal leads, and outgoing from the wire to the leads. b) The high-momentum interaction process of the chiral currents illustrated on the electrons' dispersion curve. c) The tunneling and backscattering of the fractional wire's gapped  and ballistic modes, respectively.}
\end{figure}

However, here we show that the shot-noise of a strongly correlated helical wire with fractional conductance, is disparate. Helical wires consist of counter propagating 1D modes of opposite spin. They were realized in semiconductors either with spin orbit coupling under a magnetic field~\cite{dgg10}, or possibly due to internal magnetic ordering~\cite{Zumbuhl14}. Sufficiently strong electron electron  interactions in such a system are predicted to stabilize fractional helical states with fractional conductance~\cite{OregSelaStern}. Note that the fractional helical wire considered here is a special case of a general class of partially gapped 1D conductors in which the tunneling current flows parallel to a ballistic channel; see Fig.~\ref{fg:1}(c). We find, that in such systems Coulomb interaction inside the wire and the existence of the second ballistic channel affect the Fano-factor $q^*$ in two nontrivial manners.
(i) The tunneling charge may drag additional nonquantized charges from the ballistic channel, yielding an interaction dependent local tunneling charge $q_{\rm{tun}}$. (ii) A partial reflection of the tunneling charge results from a gradual screening of the Coulomb interaction at the source and drain leads.

In this work we provide a positive answer to the question of whether there is a universal information contained in the low-frequency components of the shot-noise of the aforementioned partially gapped 1D conductor. We further show that the non-universal, interaction dependent local properties of the system can be extracted from the shot-noise at finite frequency.

We study a model equivalent to a two-wire version of the Kane-Mukhopadhyay-Lubensky anisotropic tunnel coupled wires formulation of the FQH effect~\cite{Kane02}, which reproduces the Laughlin fractional QPs and gapless fractional edge states.
The fractional wire is connected to noninteracting normal leads; see Fig.~\ref{fg:1}(a).
It was found~\cite{OregSelaStern,Meng} that at filling factor $\nu = \frac{1}{2n+1}$ with integer $n$, when a large energy gap forbids QP tunneling, the wire's conductance is $G_\nu=\frac{2 \nu^2}{1 +\nu^2}\frac{e^2}{h}$ (\emph{e.g.} $G_{\frac{1}{3}}=\frac{1}{5}\frac{e^2}{h}$). This differs from the 2D FQH conductance of $\nu \frac{e^2}{h}$.

We first show that in the absence of tunneling, this fractional conductance contains no shot-noise. This contrasts the partition noise $S\propto\mathcal{T}(1-\mathcal{T})$~\cite{Blanter} induced by noninteracting electrons tunneling through a barrier with a transmission probability $\mathcal{T}$. Next, when considering the effect of tunneling events, we find this system to exhibit a remarkable \emph{universal} shot-noise.
The Fano-factor $q^*$, though different than the tunneling charge, is universal and interaction independent. For clean wires, we find it to be $q^* = \frac{2 \nu}{1 +\nu^2}e$ (\emph{e.g.} $q^* = \frac{3}{5}e$ for $\nu = \frac{1}{3}$). Furthermore, effects of disorder in the wire likewise produce a universal Fano-factor.
This universality, in our fractional partially gapped system, is rooted in the charge conservation of the chiral modes in presence of Coloumb interaction. This is analogous to the conductance quantization of quantum wires connected to noninteracting leads~\cite{2probe4probe}.

The non-universal interaction dependent local tunneling charge $q_{\rm{tun}}$, can be extracted from the shot-noise at finite frequencies $\omega \thicksim \frac{v_F}{L}$, where $L$ is the length of the wire and $v_F$ is the Fermi velocity. A similar suggestion to use finite frequency noise was made~\cite{BERGOREG} to measure the charge fractionalization~\cite{Pham}  in Luttinger liquids.

\paragraph{The Model \--}
We study an interacting quantum wire of length $L$, adiabatically connected to non-interacting normal leads, containing both a Rashba spin-orbit (SO) coupling and a Zeeman field.
The model was introduced in Ref.~\cite{OregSelaStern} and we herein recapitulate its crucial ingredients. The SO coupling horizontally shifts the electrons' dispersion relations by $\pm k_\mathrm{SO}$ according to their spin, thus creating four Fermi points. These correspond to the left and right moving electrons with either spin up or spin down, $\psi^L_\uparrow , \psi^R_\uparrow , \psi^L_\downarrow , \psi^R_\downarrow$, denoted $\psi_1 , \psi_2 , \psi_3 , \psi_4$; see Fig.~\ref{fg:1}(b).

Near the Fermi surface one can consider both low-momentum and high-momentum physical processes involving scatterings amidst the Fermi points. The former are the density-density processes, handled within the Luttinger-liquid (LL) formalism~\cite{Giamarchi} as a free bosonic Hamiltonian $\mathcal{H}_0$; see Eq.~(\ref{eq:H0}). The latter involve $2k_F$ interactions originating from multi-electron processes described by $\mathcal{H}_{\mathrm{int}}$. The total Hamiltonian is accordingly.
\begin{equation}\label{H01}
\mathcal{H}=\mathcal{H}_0+\mathcal{H}_{\mathrm{int}}.
\end{equation}
In clean wires with SO coupling, high-momentum scattering between spin up and spin down can be generated by the magnetic field and by low momentum interactions, and open partial gaps near certain electronic densities~\cite{OregSelaStern}.
As pointed out by Kane \emph{et al.}~\cite{Kane02}, the multiparticle processes shown in Fig.~\ref{fg:1}(b),
described by the Hamiltonian
\begin{equation}\label{}
\mathcal{H}_{\mathrm{int}} \thicksim (\psi_1^\dag\psi_2)^n\psi_3^\dag\psi_2(\psi_3^\dag\psi_4)^n + \mathrm{h.c.},
\end{equation}
 may open a partial gap
at filling factors $\nu \equiv \frac{k_F}{k_\mathrm{SO}}=\frac{1}{2n+1}$.

Treatment of this interaction is done within the bosonization formalism~\cite{Giamarchi}. This is done by introducing four bosonic fields $\phi_i$, such that $\psi_i \thicksim e^{i \varphi_i}$, with their associated densities $\rho_i=(-1)^i \frac{1}{2\pi}\partial_x \varphi_i$ and currents $j_i=(-1)^{i+1} \frac{1}{2\pi}\partial_t \varphi_i$. Thus, at any point in the wire or the leads, the total current is $I = e \sum_i j_i$.
The interaction may now be expressed in its bosonized form
\begin{equation}\label{}
  \mathcal{H}_{\mathrm{int}}\thicksim\cos(n\varphi_1-(n+1)\varphi_2+(n+1)\varphi_3-n\varphi_4).
\end{equation}
Following Ref.~\cite{Kane02}, we introduce new chiral bosons,
\begin{equation}\label{BB}
  \begin{split}
    \left(\begin{array}{c}
       \phi^L_\mathrm{bal} \\
       \phi^R_\mathrm{gap}
     \end{array}
    \right)&\equiv    \sqrt{\nu}\left(\begin{array}{cc}
       n+1 & -n \\
       -n & n+1
     \end{array}
    \right)\left(\begin{array}{c}
       \varphi_1 \\
       \varphi_2
     \end{array}
    \right), \\
    \left(\begin{array}{c}
       \phi^L_\mathrm{gap} \\
       \phi^R_\mathrm{bal}
     \end{array}
    \right)&\equiv    \sqrt{\nu}\left(\begin{array}{cc}
       n+1 & -n \\
       -n & n+1
     \end{array}
    \right)\left(\begin{array}{c}
       \varphi_3 \\
       \varphi_4
     \end{array}
    \right),
  \end{split}
\end{equation}
in terms of which the cosine interaction becomes a simple backscattering operator, $\cos ( \frac{\phi^R_{{\rm{gap}}}-\phi^L_{{\rm{gap}}}}{\sqrt{\nu}} )$. By defining the canonically conjugate pairs $\theta_\mathrm{gap}\pm\phi_\mathrm{gap}\equiv\phi^{L/R}_\mathrm{gap}$ and $\theta_\mathrm{bal}\pm\phi_\mathrm{bal}\equiv\phi^{L/R}_\mathrm{bal}$, the interaction term takes the simpler form
\begin{equation}\label{eq:int}
  \mathcal{H}_{\mathrm{int}} = \int dx \left[g(x)\cos \left( \frac{2\phi_{{\rm{gap}}}}{\sqrt{\nu}} \right)\right],
\end{equation}
where the effective high-momentum interaction strength $g(x)$ adiabatically changes from $0$ at the leads to $g$ inside the wire. When the scaling dimension $\Delta$ of this interaction is smaller than two, an energy-gap $E_\mathrm{gap}$ is created in the $\phi_{{\rm{gap}}}$ mode, while the ballistic mode $\phi_{{\rm{bal}}}$ remains gapless; see Fig.~\ref{fg:1}(c).

The low-momentum Hamiltonian $\mathcal{H}_0$ must be invariant under the switching of both chirality $L\leftrightarrow R$ and spin $\uparrow\leftrightarrow\downarrow$. The $\phi \equiv (\phi_{{\rm{gap}}},\phi_{{\rm{bal}}})$ fields are odd under this transformation ($\phi \rightarrow-\phi$) while the $\theta \equiv (\theta_{{\rm{gap}}},\theta_{{\rm{bal}}})$ fields are even ($\theta \rightarrow+\theta$). The most general low-momentum Hamiltonian must therefore take the form
\begin{equation}\label{eq:H0}
  \mathcal{H}_0 = \frac{\hbar}{2\pi}\int dx \left[\partial_x \phi^\top H_0^\phi(x)\partial_x \phi + \partial_x \theta^\top H_0^\theta(x)\partial_x \theta\right],
\end{equation}
where $H_0^\phi(x)$ and $H_0^\theta(x)$ are symmetric $2 \times 2$ matrices.
These matrices change adiabatically from the noninteracting leads to a Luttinger liquid inside the wire, which nevertheless doesn't affect our universal results.

\paragraph{Quantized Conductance \--}

We begin our analysis by rederiving the fractional conductance~\cite{OregSelaStern} and showing its universality~\cite{Meng}. At each end of the wire there are two  incoming currents (from the leads) and two outgoing currents (to the leads). We denote the incoming left-moving currents by $j_{i_L}^\mathrm{in}\equiv\left.j_{i_L}\right|_{x=+\frac{L}{2}}$ with $i_L\in\{1,3\}$, and the incoming right-moving currents by $j_{i_R}^\mathrm{in}\equiv\left.j_{i_R}\right|_{x=-\frac{L}{2}}$ with $i_R\in\{2,4\}$; see Fig.~\ref{fg:1}(a).
The incoming currents are determined by the chemical potential $\mu_{L,R}$ of the lead they emanate from, with a voltage difference $e V = \mu_L - \mu_R$,
\begin{equation}\label{eqG0}
\begin{array}{cc}
  -\langle j_{i_L}\rangle_{x=+\frac{L}{2}} = \frac{1}{2\pi\hbar} \mu_R,& \phantom{+}\langle j_{i_R}\rangle_{x=-\frac{L}{2}} = \frac{1}{2\pi\hbar} \mu_L.
\end{array}
\end{equation}
Taking the limit of increasingly large $E_\mathrm{gap}$,  the gapped mode $\phi_\mathrm{gap}$ becomes stationary $\langle\partial_t\phi_\mathrm{gap}\rangle=0$.
The ungapped modes satisfy $[\int_{-\frac {L}{2}}^{\frac L2}dx\partial_x\phi_{\mathrm{bal}}^{L/R},\mathcal{H}]=0$, resulting in the conservation of their associated currents,
\begin{equation}\label{eqC}
\langle\partial_t\phi^{L/R}_\mathrm{bal}\rangle_{x=+\frac{L}{2}}=\langle\partial_t\phi^{L/R}_\mathrm{bal}\rangle_{x=-\frac{L}{2}}.
\end{equation}
We now express the four equations (\ref{eqG0}) for the incoming currents in terms of $\langle\partial_t\phi \rangle_{x=\pm\frac{L}{2}},\langle\partial_t\theta \rangle_{x=\pm\frac{L}{2}}$ using Eq.~(\ref{BB}). We  set $\langle\partial_t\phi_\mathrm{gap}\rangle_{x=\pm\frac{L}{2}}=0$ and solve the linear equations for the remaining modes in either side of the wire. Substituting back to the original currents $\langle j_i\rangle$ we find the total electric current $\langle I\rangle$ flowing through the wire and the conductance~\cite{OregSelaStern}
\begin{equation}\label{Gn}
G_\nu\equiv\frac{\langle I\rangle}{V}=\frac{e}{V}\sum_i \langle j_i\rangle_{x=x'} = \frac{2\nu^2}{1+\nu^2}\frac{e^2}{2\pi\hbar},
\end{equation}
where $x'$ may be any point outside the wire in either of the leads. We emphasize that the Hamiltonian $\mathcal{H}_0$ does not affect this result, as was shown by Meng \emph{et. al.}~\cite{Meng} using other methods.

\paragraph{Universal Low Frequency Noise \--}

 We calculate the noise of the total current $I(t)$, which  at any point $x'$ inside the wire can be expressed, using Eq.~(\ref{BB}), as $\frac{1}{e} I(t) =  \frac{\sqrt{\nu}}{\pi} \partial_t (\phi_\mathrm{gap} + \phi_\mathrm{bal}) \equiv  j_{\rm{gap}}+ j_{\rm{bal}}$. We start the calculation by defining the fluctuations of any operator $O$ as
\begin{equation}\label{Sdf}
S_O(\omega)_{ij}=\int_{-\infty}^\infty dt e^{i\omega t}\langle\{\delta O_i(t),\delta O_j(0)\}\rangle,
\end{equation}
with $\delta O \equiv O-\langle O\rangle$. By taking the Fourier transform of the operator $O(t)$ over a duration $\tau$, one gets its spectral dependence $O(\omega)$ where $\omega=\omega_m=\frac{2\pi m}{\tau}$. Its continuous spectral power is related to the noise by the Wiener-Khintchine theorem $\frac1\tau\langle \left\{\delta O_i(\omega),\delta O_j(-\omega)\right\}\rangle\xrightarrow[\tau\rightarrow\infty]{}S_O(\omega)_{ij}$.

At low frequencies $\omega\ll\frac{v_F}{L}$ the spectral components of the ungapped modes vary slowly over time scales much longer than the propagation time through the wire. In this limit, to leading order in $\frac{\omega L}{v_F}$, one can rewrite Eq.~(\ref{eqC}) as a field operator equation,
\begin{equation}\label{eqB}
\left.\partial_t\phi^{L/R}_\mathrm{bal}(\omega)\right|_{x=+\frac{L}{2}}\simeq\left.\partial_t\phi^{L/R}_\mathrm{bal}(\omega)\right|_{x=-\frac{L}{2}}.
\end{equation}
A similar conservation argument for the gapped field, which follows from $[\int_{-\frac {L}{2}}^{\frac L2}dx\partial_x\phi_{\mathrm{gap}},\mathcal{H}]=0$, yields  $\partial_t\phi_\mathrm{gap}(\omega)|_{x=+\frac{L}{2}}\simeq\partial_t\phi_\mathrm{gap}(\omega)|_{x=-\frac{L}{2}}=\frac {\pi}{\sqrt{\nu}}j_{\mathrm{gap}}(\omega)$.
 We use this equation with Eqs.~(\ref{eqB}),(\ref{BB}) to express $I(\omega)$ in terms of the Fourier components of the incoming currents and the tunneling current,
\begin{multline}\label{joj1}
\left.I(\omega)\right|_{x=\pm\frac{L}{2}}=\left\{\nu\frac{1+\nu}{1+\nu^2}\left[j_1^\mathrm{in}(\omega)+j_4^\mathrm{in}(\omega)\right]\right.\\
\left.-\nu\frac{1-\nu}{1+\nu^2}\left[j_2^\mathrm{in}(\omega)+j_3^\mathrm{in}(\omega)\right]+\frac{2}{1+\nu^2}j_\mathrm{gap}(\omega)\right\}e.
\end{multline}
When the wire is fully gapped (\emph{i.e.} $E_\mathrm{gap}\rightarrow\infty$) we may neglect the contribution of $j_{\mathrm{gap}}$. The leading contribution to the electric current noise $S_I(\omega)$  stems from the thermal noise of the incoming fields $j^\mathrm{in}$, which is given by~\cite{Blanter} $S_{j^\mathrm{in}}(\omega)_{ij}=\delta_{ij}\frac{\omega}{2\pi}\coth(\frac{\hbar\omega}{2k_B T})$.
This and Eq.~(\ref{joj1}) can be utilized in order to derive the noise of the electric current
\begin{equation}\label{BGN0}
S_I(\omega)=2e^2\frac{\omega}{2\pi}\coth(\frac{\hbar\omega}{2k_B T})\frac{2\nu^2}{1+\nu^2}.
\end{equation}
This expression tends to the Johnson-Nyquist zero frequency result of $S_I=4k_B T G_\nu$.
As this holds in the presence of voltages $e V\gg k_B T$, it ushers the conclusion that although we have a non-integer conductance, there is no zero temperature noise.

We move on to evaluate the noise when the gap is finite and yet larger than the energy scales in the system $E_\mathrm{gap}\gg\{eV,k_B T\}$. At low temperatures $eV\gg k_B T$ the dominating contribution to $S_I(\omega)$ comes from tunneling events through the gap. Hence we define the shot-noise charge, or Fano-factor, as $q^\ast\equiv\frac{S_I(\omega=0)}{2\langle I_\mathrm{tun}\rangle}$, namely by the ratio of the zero-frequency noise to the tunneling current $\langle I_\mathrm{tun}\rangle\equiv \langle I\rangle-G_\nu V$. At a single tunneling event the combination of the gapped mode $\frac{2\phi_\mathrm{gap}}{\sqrt\nu}$ subject to the cosine potential changes by $2\pi$ from one minima to another. From the definition of $j_\mathrm{gap}$, the charge carried by it during the tunneling event is heuristically $\int dt [j_{\rm{gap}}] = \nu$.
Defining a tunneling rate $\Gamma=\frac{\langle j_{\rm{gap}} \rangle}{\nu}$, we may read the charge transfered between the leads at each such event off from Eq.~(\ref{joj1}) to be $q^* =  \frac{\langle I_{\mathrm{tun}}\rangle}{\Gamma}= \frac{2\nu}{1+\nu^2}e$. This charge indeed matches the Fano factor, as we consecutively show.

As $E_\mathrm{gap}\gg eV$, the tunneling events become scarce and independent. Therefore, at low frequencies $eV\gg\hbar\omega$, the spectrum exhibited by $j_\mathrm{gap}$ is Poissonian~\cite{schottky},   $S_{j_\mathrm{gap}}(\omega)=2 \nu \langle j_\mathrm{gap}\rangle$.
 Using Eq.~(\ref{joj1}) and the Poissonian form of $S_{j_\mathrm{gap}}$ we obtain the shot-noise charge
\begin{equation}\label{q0}
q^\ast=\frac{S_I(\omega=0)}{2\langle I_\mathrm{tun}\rangle} =
\frac{2\nu}{1+\nu^2} e.
\end{equation}
This is our main result. One may understand the fact that $q^*$ differs from the tunneling charge as follows. Upon arriving to a given lead, the tunneling charge that has tunneled is no longer an eigenstate, and gets partially reflected. The existence of the ballistic channels permits multiple reflections of this charge between the opposite leads. Importantly, due to the formation of the gap in the fractional wire, the original chiral charges are not conserved $[\int dx\partial_x\varphi_i,\mathcal{H}]\neq 0$, allowing the multiple reflections to modify the overall transmitted charge to a different value.
Remarkably, this shot-noise charge attains a universal value unaffected by low-momentum Coulomb interactions. This universality stems from the connection to the noninteracting leads much like the conductivity~\cite{2probe4probe}.

We briefly discuss the influence of a small amount of disorder. The leading perturbation of an impurity was shown to be $\mathcal{H}_\mathrm{BS}\thicksim \cos ( \frac{1-\nu}{\sqrt{\nu}}  \phi_{{\rm{bal}}}|_{x=x_\mathrm{BS}})$~\cite{OregSelaStern}, manifesting in backscatterings of the ballistic channel, see arrows in Fig.~\ref{fg:1}(c), and hence, lowering the electric current $\langle I\rangle = G_\nu V + \langle I_{{\rm{tun}}}\rangle- \langle I_{{\rm{BS}}}\rangle$. As long as this perturbation remains sufficiently small, we may treat these processes as independent Poissonian events. We model these events by introducing an impurity term in Eq.~(\ref{eqB}), with $\partial_t \theta_\mathrm{bal}(\omega)|_{x=+\frac{L}{2}}\simeq \partial_t \theta_\mathrm{bal}(\omega)|_{x=-\frac{L}{2}}+\frac{\pi}{\sqrt{\nu}}j_{\mathrm{imp}}$, with the impurity backscattering current obaying $\int dt\left[j_{\mathrm{imp}}\right]=\frac{1-\nu}{2}$. Then, following similar arguments as above, we find a universal Fano-factor, originating from the backscattering. The distinct topologies of the backscattering and tunneling operators, as seen in Fig.~\ref{fg:1}(c), impel the differing of their Fano-factors. We find the noise to be $S = 2 q^*  \langle I_{{\rm{tun}}}\rangle + 2 q^*_{\rm{{BS}}}  \langle I_{{\rm{BS}}}\rangle$, with $q^*_{{\rm{BS}}}= \frac{\nu(1-\nu)}{1+\nu^2}e$ (\emph{e.g.} $q^*_\mathrm{BS} = \frac{1}{5}e$ for $\nu = \frac{1}{3}$). Since the tunneling current exponentially decays with $\frac{L E_{\rm{gap}}}{ \hbar v_F }$, it is dominated by the backscattering contribution given a sufficiently long wire.

We stress that the fact that $q^*\neq q_{\mathrm{tun}}$ in our system is consistent with the equality of these two quantities appearing in many other systems, in particular for impurity noise in FQH bars. In the appendices we show that the tunneling charge $q_{\rm{tun}}$ and the Fano-factor $q^*$ coincide and reduce to the Laughlin quasi-particle charge in the 2D FQH limit of an array of many such wires coupled together.

\paragraph{High Frequency Noise \--}
Contrary to the zero frequency noise, the high frequency noise is affected by Coulomb interactions. It reveals short timescale processes, and is used to extract the initial tunneling charge $q_{\rm{tun}}$ inside the wire. This charge encompasses the screening cloud, formed by the interaction between the ballistic and gapped modes afore hitting the leads.

Closed form expressions for $S_I(\omega)$ and $q_{\rm{tun}}$ can be obtained for a broad family of low-momentum interactions that are simultaneously diagonalizable with the cosine perturbation Eq.~(\ref{eq:int})~\cite{remarkINTERACTION}. This is done via a canonical transformation $(\tilde\phi_{{\rm{gap}}},\tilde\phi_{{\rm{bal}}})^\top \equiv A^{-1} (\phi_{{\rm{gap}}},\phi_{{\rm{bal}}})^\top$ and $(\tilde\theta_{{\rm{gap}}},\tilde\theta_{{\rm{bal}}})^\top \equiv A^\top (\theta_{{\rm{gap}}}, \theta_{{\rm{bal}}})^\top$, such that $\tilde\phi_\mathrm{gap}\propto\phi_\mathrm{gap}$ (\emph{i.e.} $A_{12}=0$). The transformed low-momentum Hamiltonian $\mathcal{H}_0$ inside the wire is diagonalized with the gapped fields $(\tilde\phi_\mathrm{gap},\tilde\theta_\mathrm{gap})$ associated with a velocity $u_\mathrm{gap}$, and fully decoupled from the ballistic fields $(\tilde\phi_\mathrm{bal},\tilde\theta_\mathrm{bal})$ propagating with velocity $u_\mathrm{bal}\equiv u$. The interaction Hamiltonian takes the form $\mathcal{H}_{\mathrm{int}} =\int dx\left[g(x)\cos(\frac{2 A_{11}}{\sqrt{\nu}}\tilde\phi_\mathrm{gap})\right]$ with scaling dimension $\Delta = \frac{A_{11}^2}{\nu}$.

A tunneling event corresponds to a $2 \pi$ jump in the gapped field $\frac{2 A_{11}}{\sqrt{\nu}} \tilde\phi_\mathrm{gap}$, while leaving $\tilde\phi_\mathrm{bal}$ unaffected. The current at any point $x'$ inside the wire can be decomposed as $\frac{1}{e}I =\frac{ \sqrt{\nu}}{ \pi}[(A_{11}+A_{21}) \partial_t \tilde{\phi}_{\rm{gap}} +(A_{12}+A_{22})\partial_t \tilde{\phi}_{\rm{bal}} ]\equiv \tilde{j}_\mathrm{gap}+\tilde{j}_\mathrm{bal}$. Since the ballistic fields are decoupled from the gapped fields, the  total tunneling charge is $q_{\rm{tun}}=\int dt I=\int dt [e\tilde{j}_\mathrm{gap}] =\left(1+\frac{A_{21}}{A_{11}} \right)\nu e$. This procedure may be used in future works to enable exact crossover solutions either via Bethe-ansatz~\cite{TBA}, or at $\Delta=\frac{1}{2}$ via refermionization.

The ballistic chiral fields
$\tilde\phi^{L/R}_\mathrm{bal}\equiv\tilde\theta_\mathrm{bal}\pm\tilde\phi_\mathrm{bal}$ are free. Therefore an appropriate phase shift can be included to account for the propagation time between the leads,
\begin{equation}\label{ps1}
\left.\tilde\phi^{L/R}_\mathrm{bal}(\omega)\right|_{x=+\frac{L}{2}}= e^{\mp i\frac{\omega L}{u}}\left.\tilde\phi^{L/R}_\mathrm{bal}(\omega)\right|_{x=-\frac{L}{2}}.
\end{equation}
Repeating the low frequency derivation, this phase shift is used to derive a lengthy high frequency expression for $\left.I(\omega)\right|_{x=\pm\frac{L}{2}}$ in terms of $j^\mathrm{in}$ and $\tilde{j}_\mathrm{gap}$. We focus on the low-temperature noise when $E_\mathrm{gap}\gg eV\gg k_B T$. At high applied voltages $eV\gg\hbar\omega$ the spectrum exhibited by $\tilde{j}_\mathrm{gap}$ is Poissonian $S_{\tilde{j}_\mathrm{gap}}(\omega)=2 \frac{q_{{\rm{tun}}}}{e}\langle \tilde{j}_\mathrm{gap}\rangle$. This allows us to calculate the finite frequency electric current noise
\begin{equation}\label{St}
S_I(\omega)=2q^\ast\langle I_{\mathrm{tun}}\rangle\left\{\left(\frac{q_{\mathrm{tun}}}{q^\ast}\right)^2+ \frac{1-\left(\frac{q_{\mathrm{tun}}}{q^\ast}\right)^2}{1+\alpha^2\tan^2(\frac{\omega L}{2u})}\right\},
\end{equation}
where $\alpha = \frac{2 \nu}{A_{22}^2(1+\nu^2)}$. This result tends to the zero frequency universal value of $S_I=2q^\ast\langle I_{\mathrm{tun}}\rangle$.
 It oscillates as a function of $\omega$ between $q^*$ and a minimal value of
\begin{equation}\label{}
\min\limits_\omega\left\{\frac{S(\omega)}{2\langle I_{\mathrm{tun}}\rangle}\right\}=\frac{q_{\mathrm{tun}}^2}{q^\ast}.
\end{equation}
This minimum is interaction dependent and may serve as a probe for measuring $q_\mathrm{tun}$ affected by the low-momentum processes within the wire.

The essential condition for the realization of fractional wires is strong enough Coulomb interactions~\cite{OregSelaStern}. Hence, the effects predicted in this paper should be experimentally measurable. We believe that insights from this work will shed light on the interpretation of noise measurements in 2D FQH systems with  multicomponent edge structures.

\paragraph{Acknowledgments:} We acknowledge the discussions with O. Agam, E. Bettelheim, Y. Gefen, M. Goldstein, T. Meng, and Y. Oreg, A. Stern.
The work was supported by a Marie Curie CIG grant and the Israel Science Foundation (ES).

\appendix{}
\numberwithin{equation}{section}
\setcounter{equation}{0}
\numberwithin{figure}{section}
\setcounter{figure}{0}
\renewcommand{\thefigure}{A.\arabic{figure}}
\renewcommand{\theequation}{A.\arabic{equation}}
\section{Appendix A}
We give here a brief generalization of our model for the Kane-Mukhopadhyay-Lubensky~\cite{Kane02} formulation of the standard FQH effect. We treat an array of $M=2N$ tunnel coupled spinless wires subject to a magnetic field. By increasing the number of wires the 2D limit is gradually obtained, introducing a spatial separation of the edge and bulk physics. This model, hence allows us to study the crossover from our 1D results to the known 2D limit.

\begin{figure}
\includegraphics[width=\linewidth]{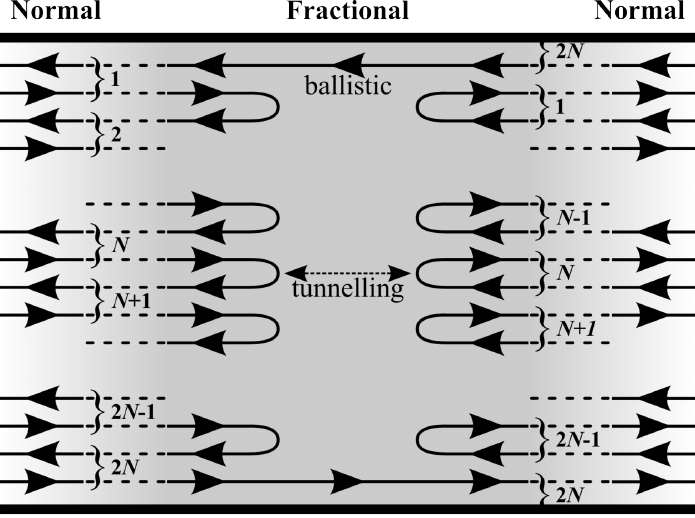}
\caption{\label{fg:A} (i) The chiral currents within the normal leads on both sides of the coupled wires' array (numbered at the left interface). (ii) The gapped modes within fractional array (numbered at the right interface). (iii) The tunneling through the equidistant mode (center). (iv) The ballistic modes propagating along the fractional edges (top and bottom).}
\end{figure}

The model is schematically depicted in Fig~\ref{fg:A}. The Hamiltonian can be split into two parts
\begin{equation}\label{}
\mathcal{H}=\mathcal{H}_0+\mathcal{H}_{\mathrm{int}}.
\end{equation}
Here $\mathcal{H}_0$ describes the effective free Hamiltonian, which includes the contribution of the low-momentum processes within the region of length $L$, and $\mathcal{H}_{\mathrm{int}}$ describes the high-momentum interactions within the same region. We focus on the most relevant processes taking place at filling factors $\nu=\frac{1}{2n+1}$ described by
\begin{equation}\label{}
\mathcal{H}_{\mathrm{int}} \thicksim \sum_{i=1}^{M-1}(\psi_i^{L\dag}\psi_i^R)^n\psi_{i+1}^{L\dag}\psi_i^R(\psi_{i+1}^{L\dag}\psi_{i+1}^R)^n + \mathrm{h.c.},
\end{equation}
where the fermionic operators $\psi^{L/R}_i$ correspond to left and right moving electrons at the Fermi surface. Within the bosonization~\cite{Giamarchi} formalism we introduce the bosonic fields $\varphi_i^{L/R}$, such that $\psi_{i}^{L/R} \thicksim e^{i \varphi_{i}^{L/R}}$, with their associated densities $\rho_i^{L/R}=\mp\frac{1}{2\pi}\partial_x \varphi_i^{L/R}$ and currents $j_i^{L/R}=\pm\frac{1}{2\pi}\partial_t \varphi_i^{L/R}$. The interaction may now be expressed in bosonized form as
\begin{equation}\label{}
  \mathcal{H}_{\mathrm{int}}\thicksim\sum_{i=1}^{M-1}
  \cos(n\varphi_i^L-(n+1)\varphi_i^R+(n+1)\varphi_{i+1}^L-n\varphi_{i+1}^R).
\end{equation}
Following Ref.~\cite{Kane02}, we make a change of variables to the canonically conjugate pairs $\left[\phi_i(x),\theta_j(x')\right]=i\frac{\pi}{2}\delta_{ij}\mathrm{sign}(x'-x)$ given by $(\phi_i,\theta_i)^\top=B_{2\times4}(\varphi_i^L,\varphi_i^R,\varphi_{i+1}^L,\varphi_{i+1}^R)^\top$ where the index $i$ is to be understood as cyclical (\emph{i.e.} $\varphi_{M+1}=\varphi_1$), with
\begin{equation}\label{}
B_{2\times4}=
\frac{\sqrt{\nu}}{2}\left(
\begin{array}{cccc}
 n  & -(n+1) & n+1  & -n  \\
-n  & n+1 & n+1  & -n
\end{array}
\right).
\end{equation}
The interaction term now takes the simple form
\begin{equation}\label{}
  \mathcal{H}_{\mathrm{int}} = \sum_{i=1}^{M-1}\int dx g(x)\cos \left(\frac{2\phi_i}{\sqrt{\nu}}\right).
\end{equation}
This interaction forms an energy-gap $E_\mathrm{gap}$ in the $\phi_1\ldots\phi_{M-1}$ modes, while maintaining the mode corresponding to the chiral edge states, $\phi_M$, as gapless.

At the ends of the coupled wires' array there are $M$ currents incoming from the leads and $M$ currents outgoing to the leads. The incoming currents are determined by the chemical potential $\mu_{L,R}$ of the lead they emanate from, with voltage difference $e V = \mu_L - \mu_R$,
\begin{equation}\label{}
\begin{array}{cc}
  -\langle j_i^L\rangle_{x=+\frac{L}{2}} = \frac{1}{2\pi\hbar}\mu_R, &
   \phantom{+}\langle j_i^R\rangle_{x=-\frac{L}{2}} = \frac{1}{2\pi\hbar}\mu_L.
\end{array}
\end{equation}
It is sufficient to focus on the zero frequency analysis. Charge conservation along the edge-states implies
\begin{equation}\label{}
\begin{split}
\left.\phi_M\right|_{x=+\frac{L}{2}}&=\left.\phi_M\right|_{x=-\frac{L}{2}},\\
\left.\theta_M\right|_{x=+\frac{L}{2}}&=\left.\theta_M\right|_{x=-\frac{L}{2}}.
\end{split}
\end{equation}
For simplicity we focus our study on bulk excitations equidistant from the edges of the array. This can be done in the symmetric case of an even number of wires $M=2N$; see Fig~\ref{fg:A}. We therefore define the current flowing through the gap by $j_\mathrm{gap}\equiv\frac{\sqrt{\nu}}{\pi}\partial_t \phi_N|_{x=0}$ while $\langle\partial_t\phi_{i\notin\{N,2N\}}\rangle=0$.

When the array is fully gapped (\emph{i.e.} $E_\mathrm{gap}\rightarrow\infty$) we set $\langle j_\mathrm{gap}\rangle=0$ and solve these equations for the modes in either side of the wire $\langle\partial_t\phi_i\rangle_{x=\pm\frac{L}{2}},\langle\partial_t\theta_i\rangle_{x=\pm\frac{L}{2}}$, allowing us to calculate the total electric current flowing through the wire and thus the conductance~\cite{OregSelaStern}
\begin{equation}\label{}
G_N\equiv\frac{\langle I\rangle}{V}= \nu\left\{1-2\left[\left(\frac{\nu+1}{\nu-1}\right)^{2N}+1\right]^{-1}\right\}\frac{e^2}{2\pi\hbar}.
\end{equation}

The low temperature noise is dominated by the contribution from $j_\mathrm{gap}$ when the gap is larger then the energy scales in the system (\emph{i.e.} $E_\mathrm{gap}\gg eV\gg k_B T$). As the tunneling events become scarce and independent, the spectrum exhibited by $j_\mathrm{gap}$ is Poissonian $S_{j_\mathrm{gap}}=2\nu \langle j_\mathrm{gap}\rangle$. This allows us to calculate the electric current noise $S_I(\omega)$. The Fano-factor charge $q^\ast\equiv\frac{S_I(\omega=0)}{2\langle I_\mathrm{tun}\rangle}$ is given by the ratio of the zero-frequency noise to the tunneling current $\langle I_\mathrm{tun}\rangle\equiv \langle I\rangle-G_N V$ which is  dominated by the contribution from $\langle j_\mathrm{gap}\rangle$ as well. Therefore a simple expression for the Fano-factor charge is derived
\begin{equation}\label{}
q^\ast_N=\left\{1-2\left[\left(\frac{\nu+1}{\nu-1}\right)^N+\left(\frac{\nu-1}{\nu+1}\right)^N\right]^{-1}\right\}\nu e.
\end{equation}
This expression tends to the Laughlin quasiparticle charge of $q^\ast_N\rightarrow \nu e$ when $N\rightarrow\infty$ and the 2D limit is approached.

\setcounter{equation}{0}
\renewcommand{\theequation}{B.\arabic{equation}}
\section{Appendix B}
The Hamiltonian inside the wire contains both the high momentum interaction $\mathcal{H}_{\mathrm{int}} = \int dx \left[g\cos \left( \frac{2\phi_{{\rm{gap}}}}{\sqrt{\nu}} \right)\right]$ and the low momentum interactions $\mathcal{H}_{0}$. By the symmetry considerations explained in the main text, the most general low momentum interactions may be represented by two real symmetric $2 \times 2$ matrices $H_0^\phi$ and $H_0^\theta$ as
\begin{equation}\label{}
  \mathcal{H}_0 = \frac{\hbar}{2\pi}\int dx \left[\partial_x \phi^\top H_0^\phi \partial_x \phi + \partial_x \theta^\top H_0^\theta \partial_x \theta\right].
\end{equation}
In general, the low momentum interactions may couple the $ \phi_{{\rm{gap}}} $ field with the $ \phi_{{\rm{bal}}} $ field, and similarly $ \theta_{{\rm{gap}}} $ with $ \theta_{{\rm{bal}}} $. Such a coupling makes the treatment of the cosine interaction non-trivial. In this appendix, we explicitly present a broad family of low-momentum Hamiltonians that are simultaneously diagonalizable with the cosine perturbation. Within this family of Hamiltonians, the effective theory decomposes into two new decoupled Luttinger liquid sectors, where the cosine interaction acts solely on one of them.

It is convenient to parameterize the two symmetric matrices by the six parameters $K_1,K_2,u_1,u_2,\gamma^\phi,\gamma^\theta$ as
\begin{equation}\label{}
\begin{split}
  H_0^\phi &= e^{-i\gamma^\phi\sigma_2}\left(
\begin{array}{cc}
 \frac{u_1 \nu}{K_1} & 0 \\
 0 & \frac{u_2 K_2}{\nu}
\end{array}
\right)e^{i\gamma^\phi\sigma_2},\\
  H_0^\theta &= e^{-i\gamma^\theta\sigma_2}\left(
\begin{array}{cc}
\frac{u_1 K_1}{\nu} & 0 \\
0 & \frac{u_2 \nu}{K_2}
\end{array}
\right)e^{i\gamma^\theta\sigma_2},
\end{split}
\end{equation}
where $\sigma_2$ is the second Pauli matrix. There are two prominent cases to note: (i) A tuning, where the gapped and ballistic modes decouple $\gamma^\phi=\gamma^\theta=0$. (ii) A standard LL Hamiltonian~\cite{Giamarchi} containing only four parameters due to symmetry under inversions of either spin or chirality. The latter Hamiltonian is attained by setting $\gamma^\phi=\gamma^\theta=\frac{\pi}{4}$, where $K_1=K_\rho,K_2=K_\sigma$ and $u_1=u_\rho,u_2=u_\sigma$ are the Luttinger parameters and velocities, of the charge and spin sectors respectively.

We construct a canonical transformation $\tilde\phi=A^{-1}\phi$ , $\tilde\theta=A^\top\theta$ that retains the shape of the cosine interaction (\emph{i.e.} $\tilde\phi_\mathrm{gap}\propto\phi_\mathrm{gap}$) while transforming the Hamiltonian to a diagonal noninteracting form.  The former condition is satisfied by using
\begin{equation}\label{}
A=\left(
\begin{array}{cc}
A_{11}&0\\
A_{21}&A_{22}\\
\end{array}
\right).
\end{equation}
In terms of $\tilde\phi$ and $\tilde\theta$, the Hamiltonian matrices read
\begin{equation}\label{}
\begin{split}
  H_0^\phi  \to \tilde{H}_0^\phi &= A^\top H_0^\phi  A, \\ H_0^\theta  \to \tilde{H}_0^\theta &= (A^{-1}) H_0^\theta  (A^{-1})^\top.
\end{split}
\end{equation}
A transformation possessing the aforementioned properties exists only when $\gamma\equiv\gamma^\phi=\gamma^\theta$ and either (i) $\gamma=0$ or (ii) $u\equiv u_1=u_2$. The former case is trivial, since $H_0^\phi$ and $H_0^\theta$ are initially diagonal. A simple observation that $\frac{1}{u^2}\tilde{H}_0^\phi\tilde{H}_0^\theta=\mathds{1}$
in the latter case, confirms the simultaneous diagonalizability of the Hamiltonian matrices. Explicitly demanding the diagonality of either of the matrices yields a condition on the ratio $\frac{A_{21}}{A_{11}}$. Recalling that the tunneling charge is $q_{\rm{tun}}=\left(1+\frac{A_{21}}{A_{11}} \right)\nu e$ we get
\begin{equation}\label{}
  q_{\rm{tun}} = \left\{1+\frac{\left( \frac{K_1}{\nu}-\frac{\nu}{K_2}  \right) \cos(\gamma) \sin(\gamma)}{\frac{K_1}{\nu} \cos^2(\gamma)+\frac{\nu}{K_2} \sin^2(\gamma)}\right\}\nu e
\end{equation}
This expression continuously interpolates between two prominent particular cases: (i) A free theory of the $\phi$ and $\theta$ fields, giving $q_\mathrm{tun} = \nu e $. (ii) A Luttinger-Liquid~\cite{Giamarchi} with velocities $u_\rho,u_\sigma$ and compressibilities $K_\rho,K_\sigma$ of the charge and spin sectors, satisfying $v_\rho = v_\sigma$, for which we obtain $q_\mathrm{tun}=\nu e\frac{2K_\rho K_\sigma}{K_\rho K_\sigma+\nu^2}$.

In order to construct the full transformation, we pursue a noninteracting form by imposing $\tilde{H}_0^\phi=\tilde{H}_0^\theta$. The transformed low-momentum Hamiltonian $\mathcal{H}_0$ inside the wire is hence both diagonalized and noninteracting. The gapped fields $(\tilde\phi_\mathrm{gap},\tilde\theta_\mathrm{gap})$ propagate with velocity $u_\mathrm{gap}=u_1$ and are fully decoupled from the ballistic fields $(\tilde\phi_\mathrm{bal},\tilde\theta_\mathrm{bal})$ propagating with velocity $u_\mathrm{bal}=u_2$. The interaction Hamiltonian takes the form
\begin{equation}\label{}
  \mathcal{H}_{\mathrm{int}} =\int dx\left[g(x)\cos(2\sqrt\Delta\tilde\phi_\mathrm{gap})\right],
\end{equation}
where the matrix element $A_{11}$ is related to the scaling dimension of the interaction via $\Delta= \frac{A_{11}^2}{\nu}=\frac{K_1\cos^2(\gamma)}{\nu^2}+\frac{\sin^2(\gamma)}{K_2}$. An explicit form of the transformation matrix is
\begin{equation}\label{}
A =  \frac{1}{\sqrt{\nu \Delta}}\left(
      \begin{array}{cc}
        \nu \Delta & 0 \\
        \left( \frac{K_1}{\nu}-\frac{\nu}{K_2}  \right) \cos(\gamma) \sin(\gamma) &  \sqrt{\frac{K_1}{K_2}}  \\
      \end{array}
    \right).
\end{equation}
To conclude, whereas the original model has six parameters characterizing the low momentum interactions, we have found a four parameter subspace, which is decomposable into one free sector and one sector with a cosine perturbation. This model is exactly solvable either via Bethe-ansatz~\cite{TBA} or upon further tuning $\Delta=\frac{1}{2}$ by means of refermionization. These may ultimately be used to explore the full crossover behavior in the system.


\begin{thebibliography}{15}
\bibitem[{rem()}]{remarkINTERACTION}
\bibinfo{note}{The explicit form of the low-momentum interactions is given in
  the appendices.}
\end{thebibliography}
\end{document}